\documentclass[twocolumn,aps,showpacs,pre,amsmath]{revtex4}
\usepackage{amssymb}
\usepackage{graphicx}

\begin{document}

\title{Aharonov-Bohm effect and giant magnetoresistance in \\ graphene nanoribbon rings}

\author{V. Hung Nguyen$^{1,2}$\footnote{E-mail: viethung.nguyen@cea.fr}, Y. M. Niquet$^1$, and P. Dollfus$^3$}
\address{$^1$L$_-$Sim, SP2M, UMR-E CEA/UJF-Grenoble 1, INAC, 38054 Grenoble, France \\ $^2$Center for Computational Physics, Institute of Physics, Vietnam Academy of Science and Technology, P.O. Box 429 Bo Ho, 10000 Hanoi, Vietnam \\ $^3$Institut d'Electronique Fondamentale, UMR8622, CNRS, Universit$\acute{e}$ Paris Sud, 91405 Orsay, France}

\begin{abstract}
 We report a numerical study on Aharonov-Bohm (AB) effect and giant magnetoresistance in rectangular rings made of graphene nanoribbons (GNRs). We show that in low energy regime where only the first subband of contact GNRs contributes to the transport, the transmission probability can be strongly modulated, i.e., almost fully suppressed, when tuning a perpendicular magnetic field. On this basis, strong AB oscillations with giant negative magnetoresistance can be achieved at room temperature. The magnetoresistance reaches thousands $\%$ in perfect GNR rings and a few hundred $\%$ with edge disordered GNRs. The design rules to observe such strong effects are also discussed. Our study hence provides guidelines for further investigations of the AB interference and to obtain high magnetoresistance in graphene devices.
\end{abstract}

\pacs{xx.xx.xx, yy.yy.yy, zz.zz.zz}

\maketitle

Graphene and its nanostructures have recently attracted a great amount of attention for both fundamental researches and device applications \cite{cast09,yazy10,novo12}. This is particularly due to its unusual electronic properties such as the linear dispersion and the chirality of carriers making graphene definitely different from conventional solid-state materials. These properties lead to many unusual transport phenomena in graphene structures such as finite minimal conductivity, Klein tunneling, or unconventional quantum Hall effect (e.g., see the review \cite{cast09}). Additionally, graphene also possesses outstanding properties such as high carrier mobility \cite{bolo08} and small spin-orbit coupling \cite{kane05} which make it very promising for applications in electronics and for use in ballistic spin transport devices. Various studies in this direction have hence been carried out (e.g., see the reviews \cite{yazy10,novo12}).

The Aharonov-Bohm (AB) oscillations \cite{ahar59} in mesoscopic rings are a phenomenon of particular interest and an elegant way to study phase coherent transport. In the presence of a perpendicular magnetic-field $B$, the phase coherent trajectories of charge carriers encircling the ring are characterized by the phase difference $\Delta \phi = 2\pi BS/\phi_0$, where $\phi_0 = h/e$ and $S$ is the area of the ring. Therefore, the transmission probability through the ring exhibits oscillations when varying the magnetic field with period $\Delta B = \phi_0/S$. The AB effect was originally observed in metal rings \cite{webb85}, and later in semiconductor heterostructures \cite{datt85}, carbon nanotubes \cite{bach99,lass07}, and topological insulators \cite{peng10}. The AB oscillations have been also explored in mesoscopic graphene rings (see the recent review \cite{sche12a}). Experimentally, clear $h/e-$AB oscillations have been observed in monolayer graphene rings \cite{russ08,huef10,smir12,rahm13}, graphene films with antidot arrays \cite{shen08}, and thin graphite crystals with columnar defects \cite{laty10}. On the theoretical side, many interesting effects have been investigated and discussed, including the valley degree of freedom typical of graphene, the influence of particular device geometries and edge symmetries, a resonant behavior with transistor applications, or the interplay between the AB effect and Klein tunneling \cite{rech07,wurm10,baha09,tluo09,sche10,zwu010,munr11,sche12b}. However, in almost all structures studied previously, the phase coherence was not as strong as expected and hence the amplitude of AB oscillations and magnetoresistance (MR) was relatively small even at low temperature. This will be discussed in more detail in this paper, on the basis of our investigations.
\begin{figure}[!t]
\centering
\includegraphics[width=3.0in]{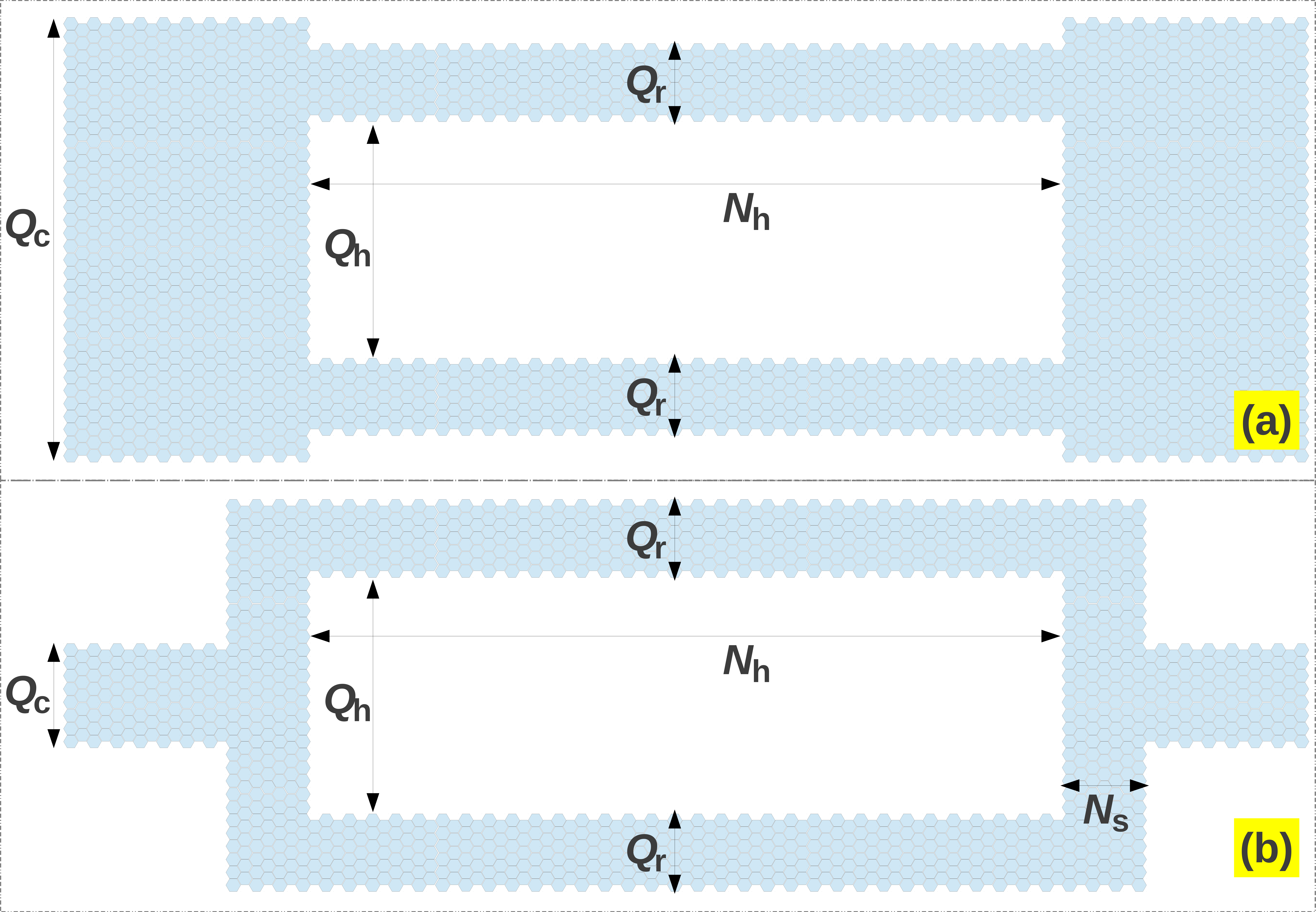}
\caption{Schematic of the graphene nanoribbon rings considered in this work. $Q_r$, $Q_c$, and $Q_h$ characterize the width of the ring, of the contact graphene nanoribbons and of the hole, respectively. $N_h$ defines the length of the hole and $N_s$ stands for the size of side nanoribbons along the transport direction.}
\label{fig_sim0}
\end{figure}

It is well known that achieving a high magnetoresistance is especially crucial for applications such as high-density data storage or magnetic sensors and actuators \cite{park02}. Hence, the investigation of this effect in graphene nanostructures with either ferromagnetic (e.g., see the review \cite{yazy10}) or non-magnetic contacts  \cite{jbai10,kuma12a,lian11,kuma12b,sing12,oost10,ribe11,mink12,upps12,zhao12} has recently been an emerging research topic. For instance, it has been experimentally reported \cite{jbai10} and theoretically demonstrated \cite{kuma12a} that a high MR of $\sim$ 50 $\%$ can be obtained at room temperature in graphene nanoribbon (GNR) devices thanks to the reduction of bandgap induced by the presence of a magnetic field. Similarly, large MR has been observed in \textit{p-i-n} GNR heterostructures \cite{lian11}. In ref. \cite{kuma12b}, a MR close to 85 $\%$ at room temperature has been achieved thanks to the orthogonality of the wavefunctions in metallic and semiconducting GNR sections. In ref. \cite{sing12}, a large ($\sim$ 50 $\%$) MR has been experimentally shown in multilayered epitaxial graphene. Additionally, the low-temperature magnetotransport has been also studied in various works \cite{oost10,ribe11,mink12,upps12,zhao12}.

In this article, we investigate the magnetotransport in the rectangular GNR rings schematized in Fig.1 and predict strong AB oscillations with a huge room-temperature MR. These strong effects are observed in the low energy regime where only the first subband of the contact GNRs carries current. Our calculations show that a negative MR of thousands $\%$ in perfect GNR rings and a few hundred $\%$ in edge disordered ones can be achieved. We also reach the conclusion that it is hard to observe such strong effects in the rings previously studied in the literature because of the multisubband contribution of contact GNRs to the transport and/or of their inhomogeneous geometries.

We use the nearest neighbor $\pi$-orbital tight binding model \cite{cast09,sche12b,hung09} to compute the electronic transport in GNR rings under a uniform perpendicular magnetic-field ($B$-field). In the presence of the $B-$field, the tight binding Hamiltonian $H$ is modified within the Peierls phase approximation \cite{mink12,peie33}. The hopping integral between nearest-neighbor atoms is hence given by $t_{nm} = -\tau_0 \exp(i\phi_{nm})$, where $\tau_0 \approx 2.7$ eV \cite{sche12b} and ${\phi_{nm}} = \frac{{2\pi }}{{{\phi _0}}}\mathop \smallint \nolimits_{{{\rm{r}}_n}}^{{{\rm{r}}_m}}$\textbf{A}(\textbf{r})\emph{d}\textbf{r}. The vector potential \textbf{A}(\textbf{r}) $= (-By,0,0)$ is related to the magnetic field \textbf{B} $= (0,0,B)$ by $\nabla\times$\textbf{A} = \textbf{B}. The charge transport through the ring is computed using an adaptive recursive Green's function method, capable of treating systems of arbitrary shape \cite{mazz12}. The linear conductance and the current are calculated using the Landauer formula
\begin{eqnarray}
 \mathcal{G}\left( B \right) &=& {G_0}\int_{ - \infty }^{ + \infty } {\mathcal{T}\left( {\varepsilon ,B} \right)\left( { - \frac{{\partial f}}{{\partial \varepsilon }}} \right)d\varepsilon }, \\
 \mathcal{I}\left( B \right) &=& \frac{{{G_0}}}{e}\int_{ - \infty }^{ + \infty } {\mathcal{T}\left( {\varepsilon ,B} \right)\left[ {{f_L}\left( \varepsilon  \right) - {f_R}\left( \varepsilon  \right)} \right]d\varepsilon },
\end{eqnarray}
where ${f_{L\left( R \right)}}\left( \varepsilon  \right) = {\left[ {1 + \exp \left( {\left( {\varepsilon  - {E_{FL\left( R \right)}}} \right)/{k_b}T} \right)} \right]^{ - 1}}$ is the left (right) Fermi distribution function with Fermi level $E_{FL(R)}$ and $G_0 = 2e^2/h$ is the quantum conductance. The transmission probability is computed as $\mathcal{T}\left( {\varepsilon ,B} \right) = Tr\left[ {{\Gamma _L}{G^r}{\Gamma _R}{G^{r\dag }}} \right]$ from the device retarded Green's function ${G^r} = {\left[ {E + i{0^ + } - H - {\Sigma _L} - {\Sigma _R}} \right]^{ - 1}}$, ${\Gamma _{L\left( R \right)}} = i\left( {{\Sigma _L} - \Sigma _L^\dag } \right)$, and the self energy ${\Sigma _{L\left( R \right)}}$ defining the left (right) contact-to-device coupling. Finally, the magnetoresistance is defined as $MR = \left[ {\mathcal{I}\left( B \right) - \mathcal{I}\left( 0 \right)} \right]/\mathcal{I}\left( B \right)$ under a finite bias or $MR = \left[ {\mathcal{G}\left( B \right) - \mathcal{G}\left( 0 \right)} \right]/\mathcal{G}\left( B \right)$ at zero bias.
\begin{figure}[!t]
\centering
\includegraphics[width=3.4in]{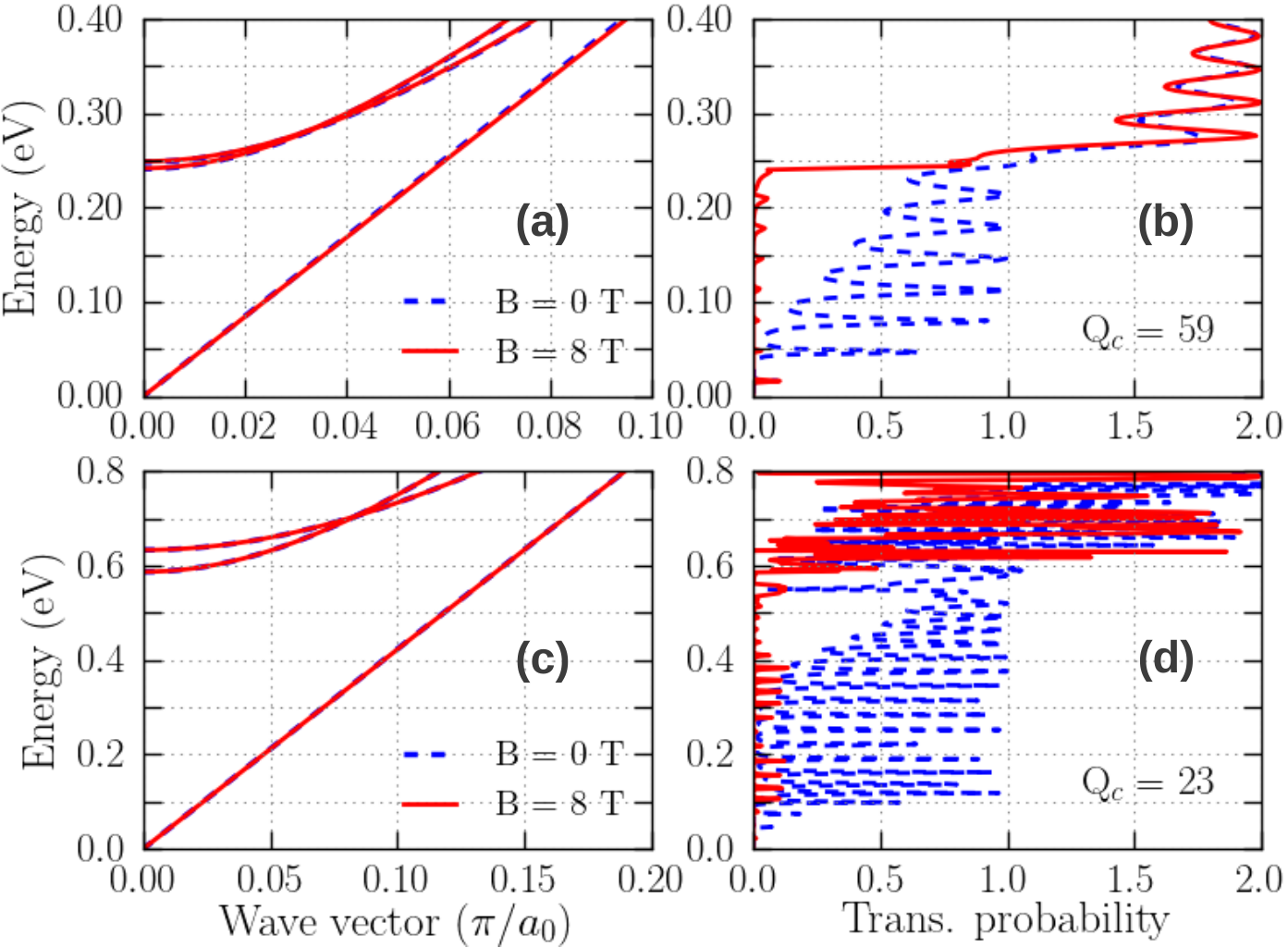}
\caption{(a,c) bandstructure of the contact GNR and (b,d) corresponding transmission probability of armchair GNR rings at $B$ = 0 and 8 T. Panels (a,b) and panels (c,d) are for the rings schematized in fig. 1(a) and fig. 1(b), respectively. The wave vector is given in units of $\pi/a_0$, with $a_0 = 3a_c$. Parameters: $Q_r = 23$, $Q_h = 13$, $N_h = 120$ and $N_s = 11$ in (c,d).}
\label{fig_sim1}
\end{figure}

Let us first investigate the properties of AB interferences in the considered rings. The ring geometry is characterized by the set of parameters of Fig. 1. The width of the GNRs ($Q_c$, $Q_r$, $Q_h$) is given in units of $a_c\sqrt{3}/2$ and $3a_c/2$ while their length ($N_h$, $N_s$) is given in units of $3a_c$ and $a_c\sqrt{3}$ in armchair and zigzag GNR rings, respectively, with $a_c = 1.42$ $\rm \AA$. In Fig. 2, we display the bandstructure of contact GNRs (left panels) and the transmission probability (right panels) of two different armchair GNR rings: Figs. 2(a,b) for the ring shown in Fig. 1(a), and Figs. 2(c,d) for the ring of Fig. 1(b). Both the contact and ring GNRs are metallic with a negligible bandgap. The presence of a $B$-field does not affect significantly the bandstructure of the contact GNRs because they are not large enough. However, an interesting phenomenon is found: due to the AB interference (shown below), the transmission probability can be strongly suppressed in the energy regime corresponding to the first subband of the contact GNRs, while the influence of the $B-$field is weak at higher energies. We suggest that these features can be understood as follows. At low energy, the contacts inject a pure state of incoming wave into the ring and the AB interference can be perfectly achieved. At high energies, i.e. when several subbands can carry electrons, the incoming wave is no longer a pure state and hence the AB interference can not take place properly. We find that these features can be reproduced in all rings with different parameters, regardless of the metallic/semiconducting or armchair/zigzag character of the GNRs (see below). In the general case, the energy regime where a strong AB interference takes place is determined by $\vert E \vert \in E_{sAB} \equiv \left[E_{e1},E_{e2}\right]$, in which $E_{e2}$ is the second band-egde of the contact GNRs and $E_{e1}$ is the lowest of the first band-edges of the contact and ring GNRs. The best option for achieving large $E_{sAB}$ and thus strong AB effects is to use semimetal GNRs and narrow contacts. The phenomenon observed above is a key-point that motivates us to investigate the AB interference and the possibility to obtain high magnetoresistance in the considered rings. Regarding the rings schematized in Fig. 1(b), we focus here on the cases of $Q_c > Q_h$ (at variance with the studies in \cite{zwu010,munr11}) to observe a strong MR effect, as discussed later.
\begin{figure}[!t]
\centering
\includegraphics[width=3.4in]{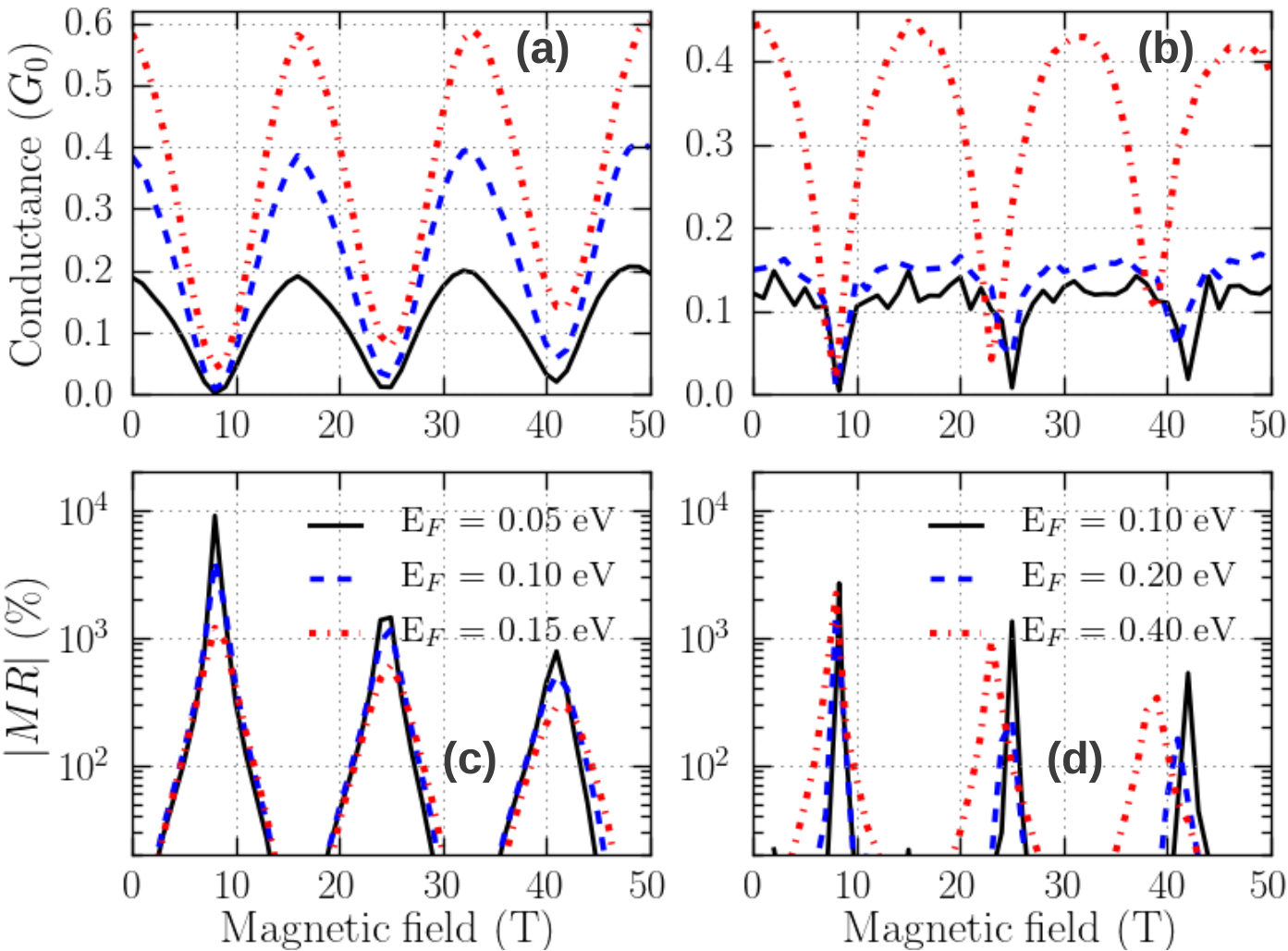}
\caption{(a,b) conductance and (c,d) corresponding magnetoresistance in armchair GNR rings as a function of $B-$field for different Fermi energies. Panels (a,c) and panels (b,d) are for the rings studied in fig. 2(b) and fig. 2(d), respectively.}
\label{fig_sim2}
\end{figure}

To clarify how strong the AB effect can be, we plot in Fig. 3 the conductance and the corresponding MR as a function of $B-$field for different Fermi energies in the two rings studied above. Note that in what follows, all transport quantities are calculated at room temperature. It is shown that (i) the conductance exhibits clear AB oscillations (see Figs. 3(a,b)), the period of which matches well the expression $\Delta B = \phi_0/S$, i.e., $\Delta B \approx$ 16 T for $S \approx 258$ nm$^2$; (ii) a giant negative MR of about a few thousand percents (see Figs. 3(c,d)) can be achieved. Here, $S$ is determined as $S = (S_{inn}+S_{out})/2$ from the inner $S_{inn}$ and outer $S_{out}$ surface areas. For completeness, we display in Fig. 4 the data obtained in rings made of zigzag GNRs. Similarly to the armchair cases, strong AB oscillations with giant MR are obtained. However, the transport at low energy takes place in the zigzag rings mainly via the edge localized states in the GNR arms, which weakens the confinement effects \cite{zwu010,munr11}. Hence, the transmission probability (and conductance peaks) in the phase coherent cases is higher than in the armchair rings. This leads to AB oscillations of large amplitude (see in Figs. 4(a,b)), so that an extremely strong MR of up to a few ten thousand percents (see Fig. 4(d)) can even be achieved for the ring of Fig. 1(b) with a large $E_{sAB}$. A similar giant modulation of the conductance, which was explained by the presence of field-induced energy gap, has been also explored experimentally in ballistic carbon nanotubes \cite{lass07}.
\begin{figure}[!t]
\centering
\includegraphics[width=3.4in]{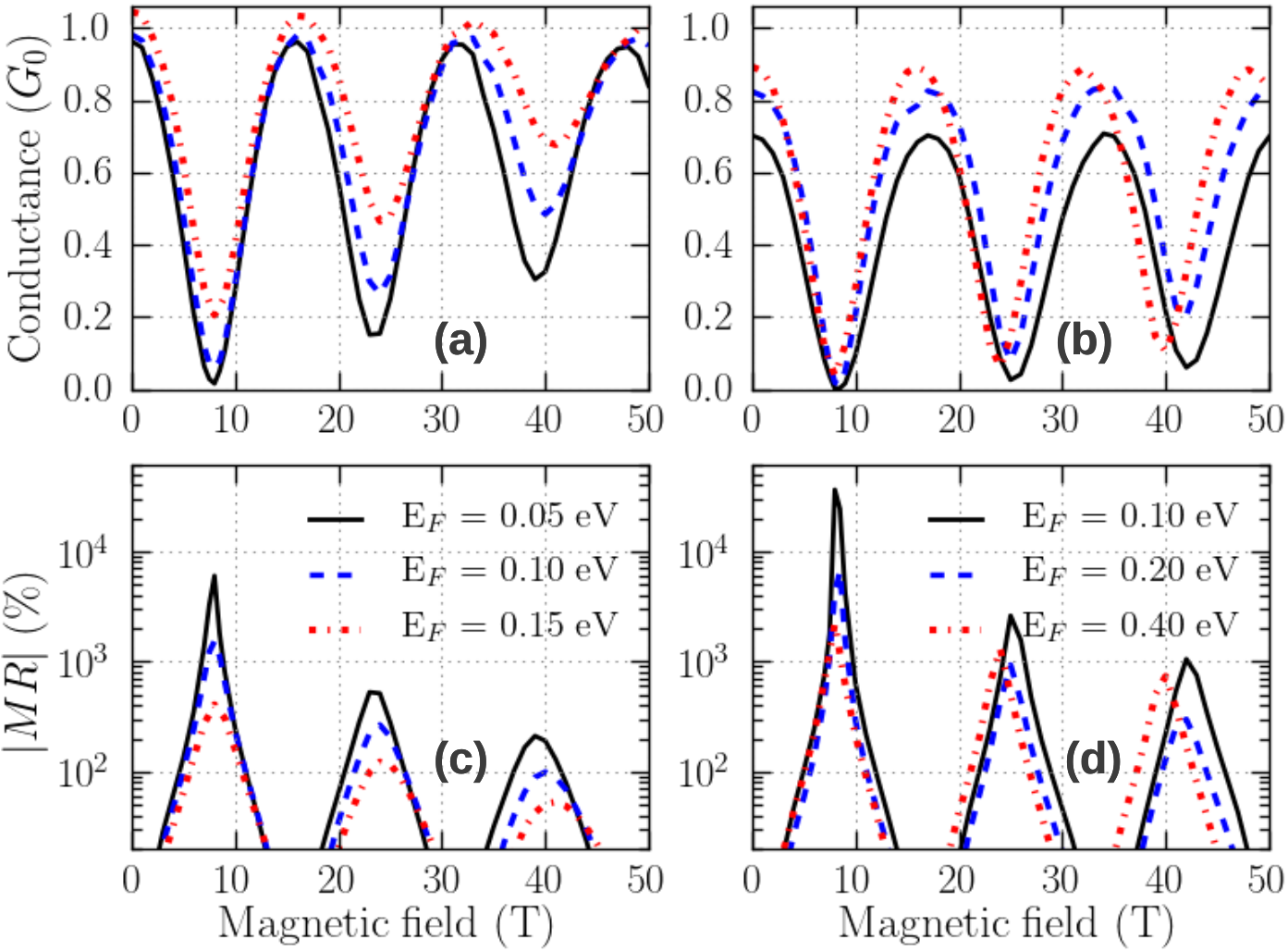}
\caption{(a,b) conductance and (c,d) corresponding magnetoresistance in zigzag GNR rings as a function of $B-$field for different Fermi energies. Panels (a,c) and panels (b,d) are for the rings having the same geometry as in fig. 1(a) and fig. 1(b), respectively. Parameters: $Q_r = 22$, $Q_h = 14$, $N_h = 120$, $Q_c = 58$ in (a,c) and $26$ in (b,d), and $N_s = 11$ in (b,d).}
\label{fig_sim3}
\end{figure}

Next, we explore the $I-V$ characteristics of the considered rings. In Fig. 5, we display the $I-V$ curves of the four rings studied in Figs. 3 and 4. Interestingly, a giant MR can still be obtained in the finite bias regime. At variance with the devices studied in \cite{jbai10,kuma12a,lian11,kuma12b} where the conduction gap at low bias is reduced, the structures considered here can switch from metallic to semiconducting behavior with an enhancement of the conduction gap when applying a $B-$field. As discussed above, the regime $E_{sAB}$ in which the strong AB interference is observed is the energy regime where only a single band of contact GNRs is active. This regime is strongly dependent on the electronic structure of the contact GNRs (see in Fig. 2), i.e., on the energy spacing between their first and second bandedges, which is, in principle, enlarged when decreasing the GNR width. The results in Fig. 5 hence show that the change in the width of the contact GNRs is a way to tune the value of $E_{sAB}$ and the bias window where the conduction gap takes place. Additionally, the possibility of enlarging $E_{sAB}$ (by reducing the width of the contact GNRs) is an advantage of the ring of Fig. 1(b) compared with that of Fig. 1(a). These results are very promising for the design of magnetic transistors as proposed in \cite{jbai10,kuma12b}. Moreover, a specific feature, the appearance of low (even negative) differential conductance at high bias, is observed in the zigzag rings. This feature (similarly, see the detailed discussion in \cite{vndo10}) can be briefly explained as follows. On the one hand, the transmission between the subbands of different parity (in particular, between the first conduction band and the first valence band at high bias) is forbidden in the GNR structures with an even number of zigzag lines (i.e., the parity selective tunneling \cite{wang08}). On the other hand, because of the change in carrier wave vector, the transmission through a step-like potential is generally smaller between different subbands (at high bias) than between same subbands \cite{vndo10}, regardless of their parity. The parity selection rule results in a conduction gap, which, together with the low transmission between different subbands mentioned above, makes the current at high bias smaller than that at low bias, i.e., the negative differential conductance (NDC) as observed in Figs. 5(c,d) and in \cite{wang08,vndo10}. The parity selection rule does not apply in the zigzag rings with an odd number of zigzag lines and their $I-V$ characteristics (not displayed here) hence do not show NDC behavior (nor do armchair GNR rings).
\begin{figure}[!t]
\centering
\includegraphics[width=3.4in]{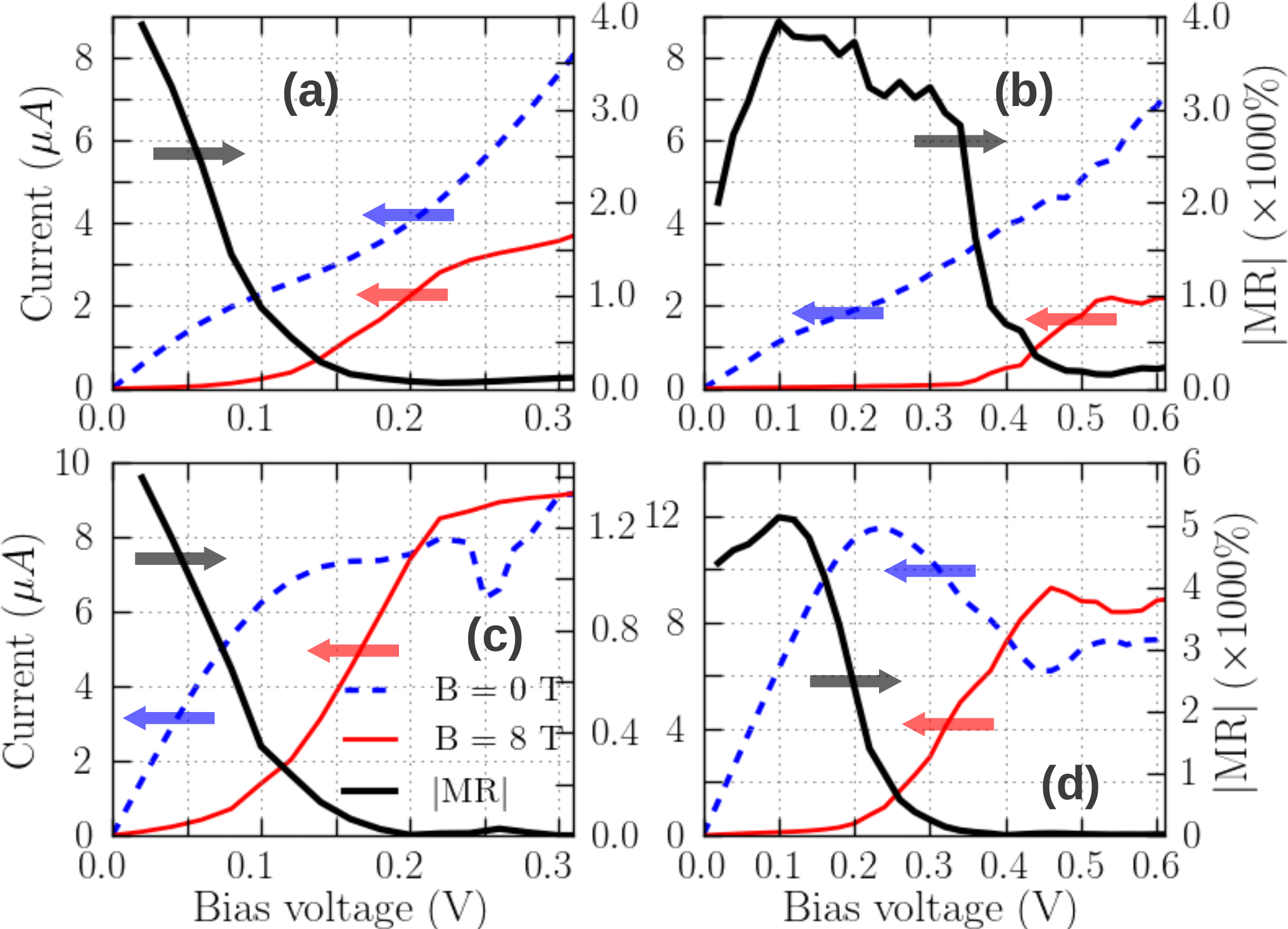}
\caption{$I-V$ characteristics (see the left axis) at $B$ = 0 and 8 T and corresponding magnetoresistance (see the right axis) of different GNR rings. Panels (a,b,c,d) correspond to the rings studied in fig. 3(a), fig. 3(b), fig. 4(a), fig. 4(b), respectively. The Fermi energy is $E_F$ = 0.1 eV in (a,c) and $E_F$ = 0.2 eV in (b,d).}
\label{fig_sim4}
\end{figure}

Though high $B$-field (i.e., from a few to a few tens Teslas) measurements have been realized in some experiments \cite{jbai10,oost10,mink12}, it is worth noting that strong AB oscillations can still be achieved at low $B-$field when increasing the ring size. To demonstrate this point, we display in Fig. 6 the conductance obtained at $B = 0$ T and in the first valley of the $\mathcal{G}\left( B \right)$-curves and the corresponding MR-peak as a function of the ring length $L_r$. It is shown that the period $\Delta B$ (and the $B-$field value of the first conductance valley) is indeed reduced proportionally to $1/L_r$, so that AB oscillations can be observed at low $B-$field when the ring is long enough. Especially, the amplitude of the MR-peaks even increases when increasing $L_r$. This feature can be understood as follows. When increasing the $B-$field, the incoming and outgoing waves are spatially separated to the ribbon edges \cite{kuma12b}, as the edge states in the quantum Hall effect. This weakens the AB interference, so that the conductance in the $\mathcal{G}\left( B \right)$-valleys can not be completely suppressed at high $B-$fields, an effect similar to the wavefunction distortion discussed in \cite{kuma12b}. This phenomenon is also evidenced by the results displayed in Figs. 3 and 4, i.e., the conductance valley increases with $B-$field. When increasing $L_r$, while $\mathcal{G}\left( 0 \right)$ is not strongly affected, the conductance valleys are observed at lower $B-$field and hence the stronger AB interference results in smaller conductance values. As a consequence, higher MR-peaks are achieved for longer $L_r$. Because the edge states are more strongly pronounced, this effect is even more significant in the zigzag rings than in the armchair ones studied in Fig. 6. However, we also notice that the increase of MR as a function of $L_r$ should be valid only in the ballistic approximation and is limited to $L_r-$values close to the graphene mean free-path, i.e., possibly about and even larger than 1 $\mu m$ in graphene on hexagonal boron nitride substrate \cite{mayo11}.
\begin{figure}[!t]
\centering
\includegraphics[width=2.8in]{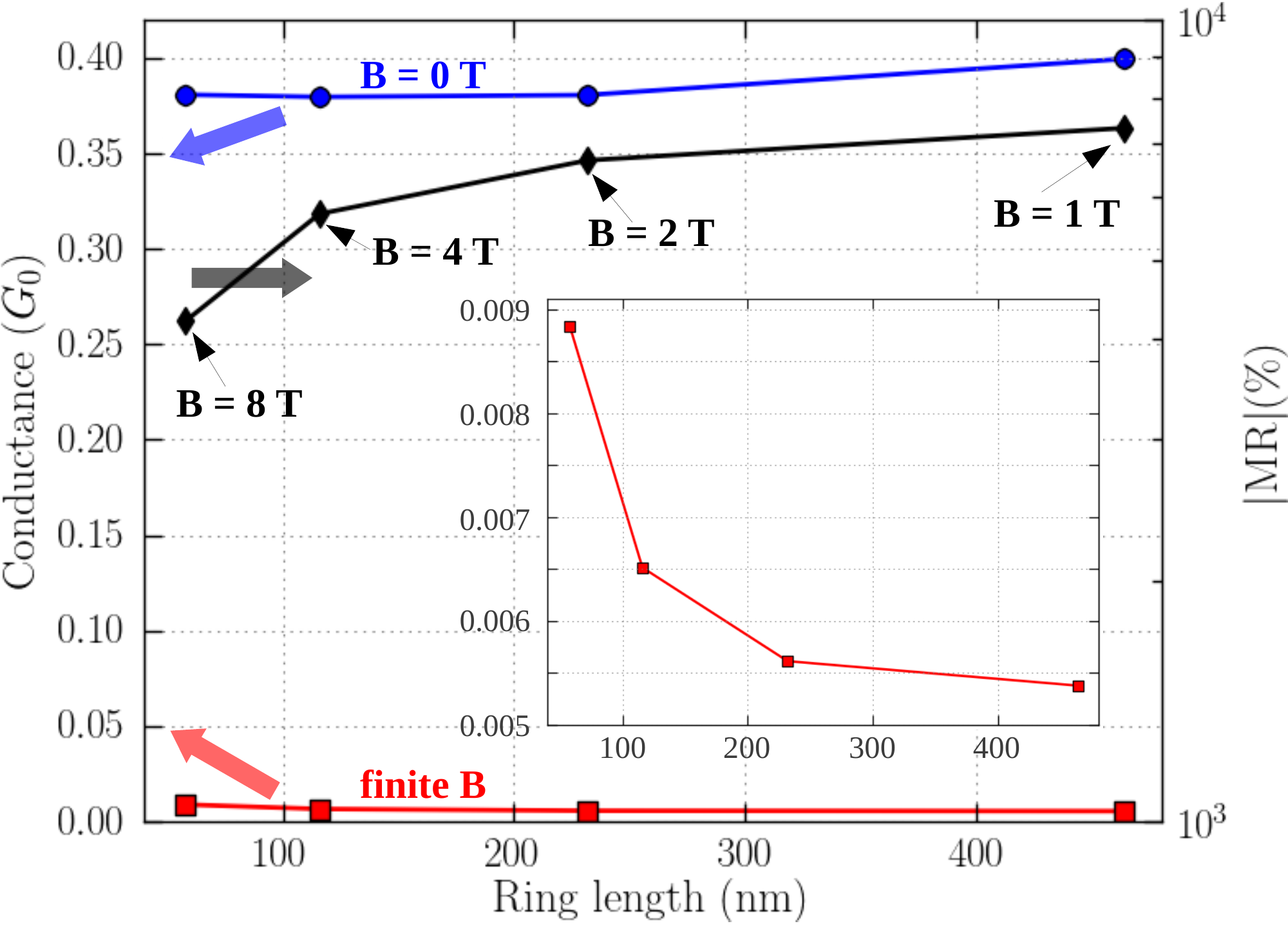}
\caption{Evolution of conductance at B = 0 T and in the first valley of $\mathcal{G}\left( B \right)$-curves (see the left axis) and of the corresponding MR-peak (see the right axis) as a function of the ring length. The inset shows a zoomed image of the conductance in the first valley. The data were computed in the rings similar to that studied in fig. 3(a) and for $E_F = 0.1$ eV.}
\label{fig_sim5}
\end{figure}

One more important point to consider is the effects of edge disorder, which are known to degrade the performance of most GNR devices. In Fig. 7, we display the MR as a function of $B-$field with different disorder configurations in the two rings studied in Figs. 3(a) and 4(a). The edge disorder is simply generated by randomly removing the edge atoms with a probability $P_D$. Indeed, the disorder strongly affects the results, i.e., it is hard to completely switch off the current with the AB interference and hence the MR amplitude is much reduced. This is due to the fact that on the one hand, the electronic properties of the system are strongly modified by the disorder and, on the other hand, complex phase shifts are induced by the scattering of wavefunctions by the defects along the ring arms. These two effects totally weaken the AB interference. However, it is worth noting that in the disordered rings studied here, a large MR of a few hundred percents can still be achieved. Moreover, besides the top-down techniques successfully used to fabricate narrow GNRs at the nanometer scale, ultra-narrow $<$5nm GNRs have been recently realized using surface-assisted bottom-up techniques \cite{jcai10,huan12,ruff12,blan12}, with atomically precise control of their topology and width. These techniques not only allow for the fabrication of ultra-narrow GNRs but also give access to GNR heterostructrures \cite{blan12}. Based on this, one can optimistically expect that the fabrication of our considered rings can be achieved soon without or with a weak edge disorder.

Additionally, besides the edge disorder, the other disorders induced by the substrate (e.g., SiO$_2$, SiC, or high $\kappa$ insulators) of graphene devices can also affect the AB interference. Fortunately, it has been recently shown that the hexagonal boron nitride (h-BN) substrate \cite{dean10,jxue11,zome11} can help to solve these issues and achieve the instrinsic properties of graphene. This is due to the fact that the surface of h-BN is flat, with a low density of charged impurities, does not have dangling bonds and is relatively inert \cite{dean10}. Ballistic transport is hence possible in a long distance, i.e., over 1 $\mu m$  \cite{mayo11}. This is certainly a good option for graphene devices able to approach our predictions.

Now, we would like to discuss the reasons (besides the disorder effects) why it has been hard to obtain the strong AB inteference in the structures previously studied in the literature. First of all, in the rings made of large contact GNRs, the AB interference is relatively weak because of the multisubband contribution to the transport as shown above. Additionally, with respect to the rectangular rings studied here, the other systems always suffer from strong inhomogeneities along the transport trajectories, as a consequence of the irregular edges of the GNRs in the circular rings \cite{russ08,huef10,smir12,rahm13,rech07,wurm10}, or of the mixing of different GNR sections in the other geometries \cite{baha09,tluo09,zwu010,munr11}. The inhomogeneities along the ring arms can result in complex phase shifts and hence weaken the AB interference, similarly to the disorder effects discussed above. These two reasons can explain the small MR obtained in the literature, compared to the strong effect observed here. On this basis, we note that in spite of having a large $E_{sAB}$, the ring geometry of Fig. 1(b) requires a careful design. Actually, there exists a mixing (with significant fractions) of zigzag and armchair GNRs in this ring if the side GNRs are too long. Therefore, the AB interfence can not take place properly for too large $Q_h/Q_c$ ratio and/or short $N_h$. In that case, only the resonant tunneling effect due to the ring geometry is significantly pronounced as reported in \cite{zwu010,munr11}. The condition $Q_c \gtrsim Q_h$ is hence mandatory to guarantee the strong MR effect.
\begin{figure}[!t]
\centering
\includegraphics[width=3.0in]{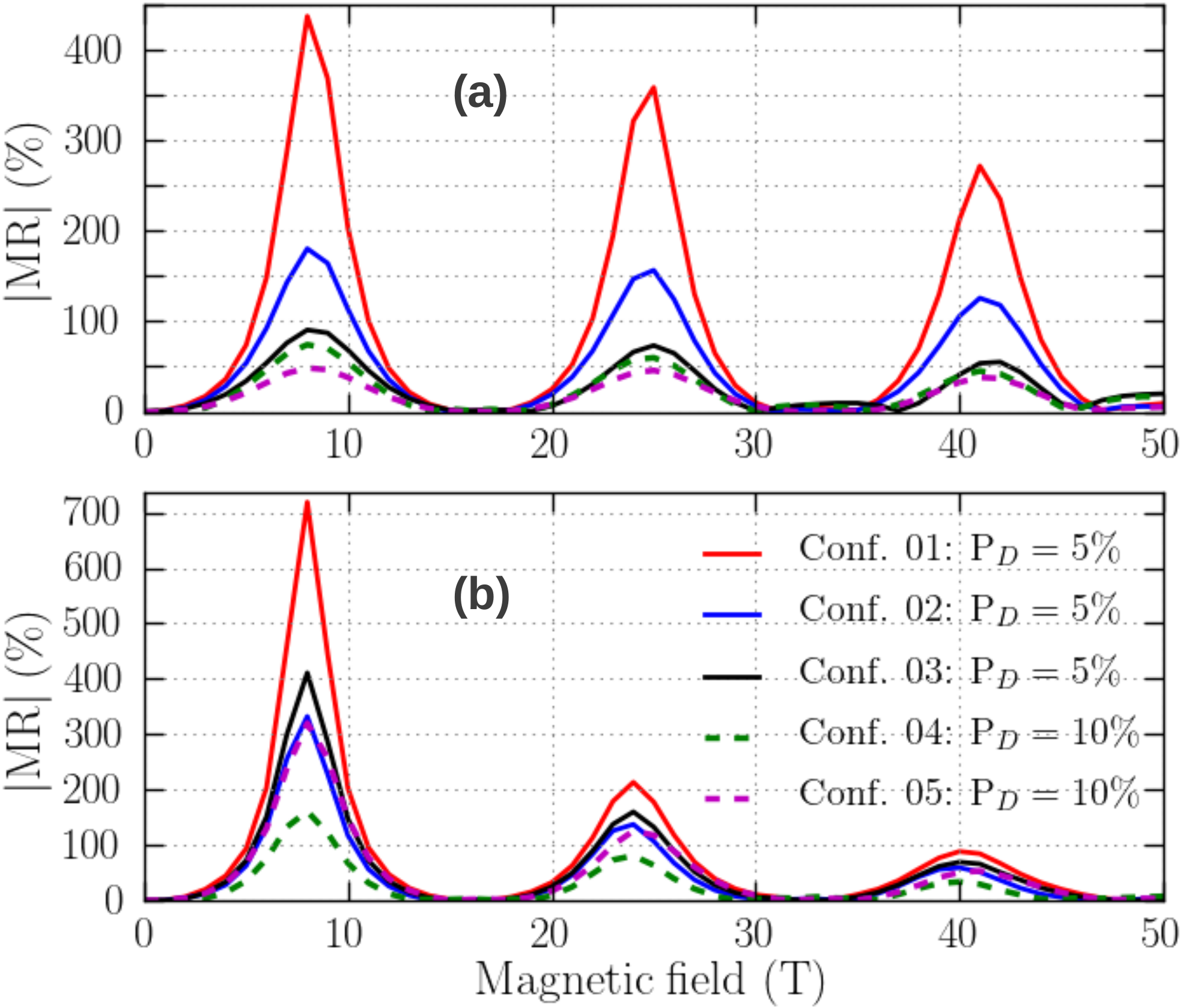}
\caption{Magnetoresistance at zero bias as a function of $B-$field with different edge disorder configurations. Panels (a) and (b) correspond to the rings studied in fig. 3(a) and fig. 4(a), respectively. The Fermi energy is $E_F$ = 0.1 eV.}
\label{fig_sim6}
\end{figure}

Finally, we also have some remarks regarding another factor, which may have an influence on our results. Since our calculations were based on the single particle theory, many body effects may affect quantitatively the results obtained in the zigzag rings, especially in ultra-narrow GNRs \cite{gruj13}. These effects can give rise to a small bandgap and to edge localized states with an antiferromagnetic interedge superexchange interaction in the zigzag GNRs. The influence of such phenomena on the AB interference is certainly a valuable objective of further works. However, on the one hand, both the bandgap and interedge coupling have been shown to strongly decrease when increasing the ribbon width \cite{wson06,jung09} and are hence negligible in wide enough GNRs, i.e., if the ribbon width is much larger than 26 $\rm \AA$ \cite{zhao13}. On the other hand, the strong AB effects observed are essentially dependent on the geometrical symmetry and on the homogeneity of the ring arms along the transport direction, which are not broken by the many-body effects as they can be in the cases of the disorders discussed above. On this basis, because all zigzag GNRs studied here have a width larger than $\sim$ 47 $\rm \AA$ ($Q = 22$), it can be expected that including the many-body effects would not strongly affect our results.

In summary, we have investigated the AB effect in rectangular GNR rings using numerical simulation within a tight binding model. We have shown that in low energy regime where only the first subband of contact GNRs contributes to the transport, i.e. in the case of a pure-state incoming wave, the transmission probability can be almost fully suppressed due to the AB interference. This suggests the possibility of tuning the structure from metallic to semiconducting state. Very strong AB oscillations with giant magnetoresistance (thousands $\%$ in perfect GNR rings and a few hundred $\%$ in edge disordered GNR rings) can be achieved at room temperature. The influence of different factors governing
the AB effects has been also discussed. The study hence suggests an efficient way to investigate the AB interference in graphene nanorings and could be very helpful for designing high magnetoresistance graphene devices.

One of the authors (P.D.) acknowledges the French ANR for financial support under the projects NANOSIM-GRAPHENE (ANR-09-NANO-016) and MIGRAQUEL (ANR-10-BLAN-0304). V. Hung Nguyen acknowledges the Vietnam's National Foundation for Science and Technology Development (NAFOSTED) for financial support under the project no. 103.02-2012.42.


\begin{thebibliography}{99}
\bibitem{cast09} A. H. Castro Neto, F. Guinea,  N. M. R. Peres,  K. S. Novoselov and A. K. Geim, Rev. Mod. Phys. \textbf{81}, 109 (2009).
\bibitem{yazy10} O. V. Yazyev, Rep. Prog. Phys. \textbf{73}, 056501 (2010).
\bibitem{novo12} K. S. Novoselov, V. I. Fal'ko, L. Colombo, P. R. Gellert, M. G. Schwab, and K. Kim, Nature \textbf{490}, 192 (2012).
\bibitem{bolo08} K. I. Bolotin, K. J. Sikes, Z. Jiang, M. Klima, G. Fudenberg, J. Hone, P. Kim, and H. L. Stormer, Solid State Commun. \textbf{146}, 351 (2008).
\bibitem{kane05} C. L. Kane and E. J. Mele, Phys. Rev. Lett. \textbf{95}, 226801 (2005).

\bibitem{ahar59} Y. Aharonov and D. Bohm, Phys. Rev. \textbf{115}, 485 (1959).
\bibitem{webb85} R. A. Webb, S. Washburn, C. P. Umbach, and R. B. Laibowitz, Phys. Rev. Lett. \textbf{54}, 2696 (1985).
\bibitem{datt85} S. Datta, M. R. Melloch, S. Bandyopadhyay, R. Noren, M. Vaziri, M. Miller, and R. Reifenberger, Phys. Rev. Lett. \textbf{55}, 2344 (1985).
\bibitem{bach99} A. Bachtold, C. Strunk, J.-P. Salvetat, J.-M. Bonard, L. Forro, T. Nussbaumer, and C. Schonenberger, Nature \textbf{397}, 673 (1999).
\bibitem{lass07} B. Lassagne, J-P. Cleuziou, S. Nanot, W. Escoffier, R. Avriller, S. Roche, L. Forro, B. Raquet, and J.-M Broto, Phys. Rev. Lett. \textbf{98}, 176802 (2007).
\bibitem{peng10} H. Peng, K. Lai, D. Kong, S. Meister, Y. Chen, X.-L. Qi, S.-C. Zhang, Z.-X. Shen, and Y. Cui, Nat. Mater. \textbf{9}, 225 (2010).
\bibitem{sche12a} J. Schelter, P. Recher, and B. Trauzettel, Solid State Comm. \textbf{152}, 1411 (2012).

\bibitem{russ08} S. Russo, J. B. Oostinga, D. Wehenkel, H. B. Heersche, S. S. Sobhani, L. M. K. Vandersypen, and A. F. Morpurgo, Phys. Rev. B \textbf{77}, 085413 (2008).
\bibitem{huef10} M. Huefner, F. Molitor, A. Jacobsen, A. Pioda, C. Stampfer, K. Ensslin, and T. Ihn, New J. Phys. \textbf{12}, 043054 (2010).
\bibitem{smir12}  D. Smirnov, H. Schmidt, and R. J. Haug, Appl. Phys. Lett. \textbf{100}, 203114 (2012).
\bibitem{rahm13} A. Rahman, J. W. Guikema, S. H. Lee, and N. Markovic, Phys. Rev. B \textbf{87}, 081401(R) (2013)
\bibitem{shen08} T. Shen, Y. Q. Wu, M. A. Capano, L. P. Rokhinson, L. W. Engel, and P. D. Ye, Appl. Phys. Lett. \textbf{93}, 122102 (2008).
\bibitem{laty10} Y. U. I. Latyshev, A. P. Orlov, E. G. Shustin, N. V. Isaev, W. Escoffier, P. Monceau, C. J. van der Beek, M. Konczykowski, and I. Monnet, J. Phys.: Conf. Ser. \textbf{248}, 012001 (2010).

\bibitem{rech07} P. Recher, B. Trauzettel, A. Rycerz, Ya. M. Blanter, C. W. J. Beenakker, and A. F. Morpurgo, Phys. Rev. B \textbf{76}, 235404 (2007).
\bibitem{wurm10} J. Wurm, M. Wimmer, H.U. Baranger, K. Richter, Semicond. Sci. Technol. \textbf{25}, 034003 (2010).
\bibitem{baha09} D. A. Bahamon, A. L. C. Pereira, and P. A. Schulz, Phys. Rev. B \textbf{79}, 125414 (2009).
\bibitem{tluo09} T. Luo, A. P. Iyengar, and H. A. Fertig, and L. Brey, Phys. Rev. B \textbf{80}, 165310 (2009).
\bibitem{sche10} J. Schelter, D. Bohr, and B. Trauzettel, Phys. Rev. B \textbf{81}, 195441 (2010).
\bibitem{zwu010} Z. Wu, Z. Z. Zhang, K. Chang, F. M. Peeters, Nanotechnology \textbf{21}, 185201 (2010).
\bibitem{munr11} J. Munarriz, F. Dominguez-Adame, and A. V. Malyshev, Nanotechnology \textbf{22}, 365201 (2011).
\bibitem{sche12b} J. Schelter, B. Trauzettel, and P. Recher, Phys. Rev. Lett. \textbf{108}, 106603 (2012).

\bibitem{park02} S. S. P. Parkin, "Applications of magnetic nanostructures" in \emph{Spin Dependent Transport in Magnetic nanostructures}, ed. S. Maekawa and T. Shinjo, Taylor and Francis, pp. 237-279 (2002).

\bibitem{jbai10} J. Bai, R. Cheng, F. Xiu, L. Liao, M. Wang, A. Shailos, K. L. Wang, Y. Huang, X. Duan, Nat. Nanotechnol. \textbf{5}, 655 (2010).
\bibitem{kuma12a} S. B. Kumar and J. Guo, Nanoscale \textbf{4}, 982 (2012).
\bibitem{lian11} G. Liang, S. B. Kumar, M. B. A. Jalil, and S. G. Tan, Appl. Phys. Lett. \textbf{99}, 083107 (2011).
\bibitem{kuma12b} S. B. Kumar, M. B. A. Jalil, and S. G. Tan, Appl. Phys. Lett. \textbf{101}, 183111 (2012).
\bibitem{sing12} R. S. Singh, X. Wang, W. Chen, Ariando, and A. T. S. Wee, Appl. Phys. Lett. \textbf{101}, 183105 (2012).
\bibitem{oost10} J. B. Oostinga, B. Sacepe, M. F. Craciun, and A. F. Morpurgo, Phys. Rev. B \textbf{81}, 193408 (2010).
\bibitem{ribe11} R. Ribeiro, J.-M. Poumirol, A. Cresti, W. Escoffier, M. Goiran, J.-M. Broto, S. Roche, and B. Raquet, Phys. Rev. Lett. \textbf{107}, 086601 (2011).
\bibitem{mink12} S. Minke, S. H. Jhang, J. Wurm, Y. Skourski, J. Wosnitza, C. Strunk, D. Weiss, K. Richter, and J. Eroms, Phys. Rev. B \textbf{85}, 195432 (2012).
\bibitem{upps12} A. Uppstu and A. Harju, Phys. Rev. B \textbf{86}, 201409(R) (2012).
\bibitem{zhao12} Y. Zhao, P. Cadden-Zimansky, F. Ghahari, and P. Kim, Phys. Rev. Lett. \textbf{108}, 106804 (2012).

\bibitem{hung09} V. Hung Nguyen, V. Nam Do, A. Bournel, V. Lien Nguyen and P. Dollfus, J. Appl. Phys. \textbf{106}, 053710 (2009).
\bibitem{peie33} R. E. Peierls, Z. Phys. \textbf{80}, 763 (1933).
\bibitem{mazz12} F. Mazzamuto, J. Saint-Martin, V. Hung Nguyen, C. Chassat, and P. Dollfus, J. Comput. Electron. \textbf{11}, 67 (2012).

\bibitem{vndo10} V. Nam Do and P. Dollfus, J. Appl. Phys. \textbf{107}, 063705 (2010).
\bibitem{wang08} Z. F. Wang, Q. Li, Q. W. Shi, X. Wang, J. Yang, J. G. Hou, and J. Chen, Appl. Phys. Lett. \textbf{92}, 133114 (2008).
\bibitem{mayo11} A. S. Mayorov, R. V. Gorbachev, S. V. Morozov, L. Britnell, R. Jalil, L. A. Ponomarenko, P. Blake, K. S. Novoselov, K. Watanabe, T. Taniguchi, and A. K. Geim, Nano Lett. \textbf{11}, 2396 (2011).

\bibitem{jcai10} J. Cai, P. Ruffieux, R. Jaafar, M. Bieri, T. Braun, S. Blankenburg, M. Muoth, A. P. Seitsonen, M. Saleh, X. Feng, K. Mullen, and R. Fasel, Nature \textbf{466}, 470 (2010).
\bibitem{huan12} H. Huang, D. Wei, J. Sun, S. L. Wong, P. F. Feng, A. H. Castro Neto, A. T. S. Wee, Sci. Rep. \textbf{2}, 983 (2012).
\bibitem{ruff12} P. Ruffieux, J. Cai, N. C. Plumb, L. Patthey, D. Prezzi, A. Ferretti, E. Molinari, X. Feng, K. Mullen, C. A. Pignedoli, and R. Fasel, ACS Nano \textbf{6}, 6930 (2012).
\bibitem{blan12} S. Blankenburg, J. Cai, P. Ruffieux, R. Jaafar, D. Passerone, X. Feng, K. Mullen, R. Fasel, and C. A. Pignedoli, ACS Nano \textbf{6}, 2020 (2012).

\bibitem{dean10} C. R. Dean, A. F. Young, I. Meric, C. Lee, L. Wang, S. Sorgenfrei, K. Watanabe, T. Taniguchi, P. Kim, K. L. Shepard, and J. Hone, Nature Nanotechnol. \textbf{5}, 722 (2010).
\bibitem{jxue11} J. Xue, J. Sanchez-Yamagishi, D. Bulmash, P. Jacquod, A. Deshpande, K. Watanabe, T. Taniguchi, P. Jarillo-Herrero, and B. J. LeRoy, Nature Mater. \textbf{10}, 282 (2011).
\bibitem{zome11} P. J. Zomer, S. P. Dash, N. Tombros, and B. J. van Wees, Appl. Phys. Lett. \textbf{99}, 232104 (2011).
\bibitem{gruj13} M. Grujic, M. Tadic, and F. M. Peeters, Phys. Rev. B \textbf{87}, 085434 (2013).
\bibitem{wson06} Y.-W. Son, M. L. Cohen, and S. G. Louie, Phys. Rev. Lett. \textbf{97}, 216803 (2006).
\bibitem{jung09} J. Jung, T. Pereg-Barnea, and A. H. MacDonald, Phys. Rev. Lett. \textbf{102}, 227205 (2009).
\bibitem{zhao13} J. H. Zhao, X. Q. Dai, B. Zhao, Y. W. Dai, X. Zhao, Eur. Phys. J. B \textbf{85}, 220 (2012).

\end{thebibliography}
\end{document}